# Aspect Based Sentiment Analysis to Extract Meticulous Opinion Value

Deepali Virmani, Vikrant Malhotra, Ridhi Tyagi

*Department of Information Technology*
*Bhagwan Parshuram Institute of Technology*
*PSP- 4, Dr. K.N. Katju Marg , Sector 17 Rohini,*
*New Delhi -110089*

[1]deepalivirmani@gmail.com
[2]vikrant.malhotra.1992@gmail.com
[3]ridhi.tyagi.1992@gmail.com

**ABSTRACT**
**Opinion Mining and Sentiment Analysis is a process of identifying opinions in large unstructured/structured data and then analysing polarity of those opinions. Opinion mining and sentiment analysis have found vast application in analysing online ratings, analysing product based reviews, e-governance, and managing hostile content over the internet. This paper proposes an algorithm to implement aspect level sentiment analysis. The algorithm takes input from the remarks submitted by various teachers of a student. An aspect tree is formed which has various levels and weights are assigned to each branch to identify level of aspect. Aspect value is calculated by the algorithm by means of the proposed aspect tree. Dictionary based method is implemented to evaluate the polarity of the remark. The algorithm returns the aspect value clubbed with opinion value and sentiment value which helps in concluding the summarized value of remark.**

## I. INTRODUCTION

Sentiment analysis or opinion mining can be considered as a sub-problem under Natural Language Processing [6], [13]. It aims at identifying what is the topic or aspect of the discussion and then identifying whether positive review have been mentioned or negative. We have different forms of data over the internet that is to be analysed. Data can be structured or unstructured. Structured data is easy to operate upon where as unstructured data requires complex analysis [6]. Thus the type of data we have, decides the method we choose.

We have different levels at which we perform sentiment analysis. Document level Sentiment analysis analyses a whole document. This technique is discussed in [1]. Sentence level sentiment analysis analyses a sentence at a time and then summarizes the analysis of all the sentences. This technique is discussed in [12]. Aspect level analyses the issue or entity over which opinion is given. This is the most basic level of sentiment analysis.

Sentiment words play the most important role in identifying the sentiments in a sentence or document. Such sentiment words when combined together is called sentiment lexicon [6]. Algorithms use this sentiment lexicon to analyse the data. Analysing sentiment is a complex task. Identification of objective and subjective text is difficult. Subjective sentences are one which expresses opinion about a subject or they are a person's perspective. Objective sentences are hard to analyse. Irony in sentences is another issue. Such issues decide the complexity of the algorithm that would be used for analysis.

### A. Basic Understanding of the concept

It is necessary to understand the basic terms and definitions that would be used under sentiment analysis and opinion mining. According to [6] opinion can be defined by two components i.e. the target $g$ and sentiment $s$ on the target

$$(g,s)$$

This is a very basic definition of opinion. Here target $g$ can be an entity over which opinion is given or it can be an aspect of the entity. It is very crucial to understand the concept of entity and aspect. For example
'The processing speed of laptop is really good'.
Here the laptop is the entity or the product under analysis [8]. Processing speed of laptop is the aspect of the entity over which opinion is expressed. '$s$' is the value of the sentiment which is being expressed. $s$ can be positive, negative, neutral or a numeric value. As the problem becomes complex, the definition of opinion elaborates.

$$(g,s,h,t)$$

Here opinion has four components. Components $g$ and $s$ are explained before. '$h$' is the opinion holder or the person who has given the review. With varying situation, it is possible that a person's opinion hold more value than other person [3], [12]. Sometimes the time '$t$' at which the opinion is given also plays an important role. If we do not want to consider older opinions then we must analyse the time constraint. An entity may be analysed over different aspects. We have to find out these aspects and analyse sentiment which is being expressed. If no aspect is mentioned then the sentiment is said to be GENERAL [6]. The main task of sentiment analysis is to find out what is the entity, what are the different aspects on which opinion is given and then analysing polarity of these opinions. In this survey the value of opinion holder is considered to be 1, since in an LOR system, all teachers have same priority level. Also time constraint is not as relevant. Such type of analysis is called aspect based or feature based sentiment analysis [8]. The different types of opinions are discussed next.

*B. Different types of opinions*

While analysing opinions, two types of opinions are encountered most commonly and they are regular opinions and comparative opinions [6], [13]. Regular opinions are simple opinions. They are categorized as direct and indirect opinion. This survey aims to analyse only direct opinions. Comparative opinion compares two entities at a time having similar or different aspects. Comparative opinions are difficult to analyse and require complex algorithms [6], [13].

## II. LITERATURE SURVEY

The earlier studies under the field of sentiment analysis were based on document level sentiment analysis [1]. The research aimed at classifying the whole document as positive or negative. The basic assumption in this case was that each document expresses opinion on only one entity expressed by only one opinion holder.

Sentiment classification can be done using supervised learning techniques and unsupervised learning techniques. Supervised learning techniques include text classification based on a classifier [14]. Supervised learning technique takes into account features like 'terms and their frequency', 'parts of speech', 'sentiment words and phrases', 'sentiment shifters' and so on [6]. Unsupervised learning techniques make use of fixed syntactic patterns that occur in an opinion. This technique uses POS tagging which identifies nouns, adverbs, adjectives etc in a sentence. Based on knowledge and arrangement of these words we identify the entity, aspect and the opinion [15]. Another approach under unsupervised learning technique is maintaining dictionary of sentiment words and their weights based on which opinion. This approach also takes into consideration effect of negation or sentiment shifters [3], [8]. Later sentence level sentiment analysis and aspect level sentiment analysis also emerged as field of research. In sentence level sentiment analysis we perform analysis at sentence level. Here the basic aim is to identify subjective and objective sentences. A model, naïve Bayes classifier is used for identifying subjectivity in sentences [14]. Aspect level sentiment analysis or feature based opinion mining is the core concept behind this paper. Research has been done on aspect level sentiment analysis [5] which aims to identify various product reviews available on the internet and analysing them. Thus there are 2 basic tasks involved in aspect level sentiment analysis [6] and they are, aspect extraction and aspect sentiment classification. This paper introduces a concept of aspect value which tells how much clear or specific is the opinion that is being given. This is done using the aspect tree. The concept is utilized in an application of LOR system. The sentiments on aspect are analysed by means of dictionary of sentiment words [7].

## III. PROPOSED CONCEPT

This survey aims at performing aspect level sentiment analysis. The survey not only returns an opinion value but also on which aspect that opinion is given. The implementation of the proposed concept is for LOR system. In LOR system, the objective is to get an opinion value for a particular student. Various teachers give their remarks about a student and the algorithm analyses these remarks to return an opinion as well as aspect value.

Aspect level sentiment analysis aims at identifying the aspects of the entity. LOR system identifies these aspects in teachers' remarks and evaluates an aspect value based on the aspect tree. The aspect tree stores a defined set of aspects. The students are analysed over these aspects. Such aspects can be student's academics, sports, extra-curricular, co-curricular performance, personality traits etc. These are the major aspect categories having sub-categories under them. For example, a student can be good in DBMS. So DBMS would be a sub-aspect of subjects which is further a sub-aspect of academics. This is how a tree is maintained for aspects and weight is given to each branch of the tree. Lower the height of an aspect in the aspect tree lower is the specificity of the opinion or remark given by the teacher. This can be explained with an example. Considering two remarks,

(1)'He actively participates in co-curricular activities.'
(2)'He actively participates in Anugoonj(University Fest) and Corona.(College Event)'

Here we can see that the remark (2) is more specific than remark one. Co-curricular lies at lower level in the aspect tree whereas Anugoonj and Corona are at higher level hence having an elevated aspect value. Thus the remark (2) is more specific than the remark (1).

Every aspect is represented by a node in the tree. Every branch is assigned a weight that depends on the node to which that branch points to. When a teacher gives remarks about a particular student, the remark is fetched sentence by sentence. Now, each sentence is stored as an array of words. From the tree of aspects, an aspect is searched in that sentence. When that aspect is found, the aspect value is calculated by traversing the tree. Traversal is done from the location of that aspect in the tree to the root of the tree. The weights of branches that are traversed are multiplied. This helps in identifying how much specific is that aspect which is mentioned in the remarks. It also increases or decreases the value of the remark. Thus the aspect value tells how much clear or specific the remark is.

Once the aspect value is known the sentiment about the aspect is calculated from the array of words by means of sentiment dictionary [7]. Here each word and negations in the sentence are analysed to find out the polarity of the remark. Thus the sentiment and the aspect provide the two dimensions of the opinion. Now, summarizing the aspect and sentiment in the remark, overall opinion about the student is obtained.

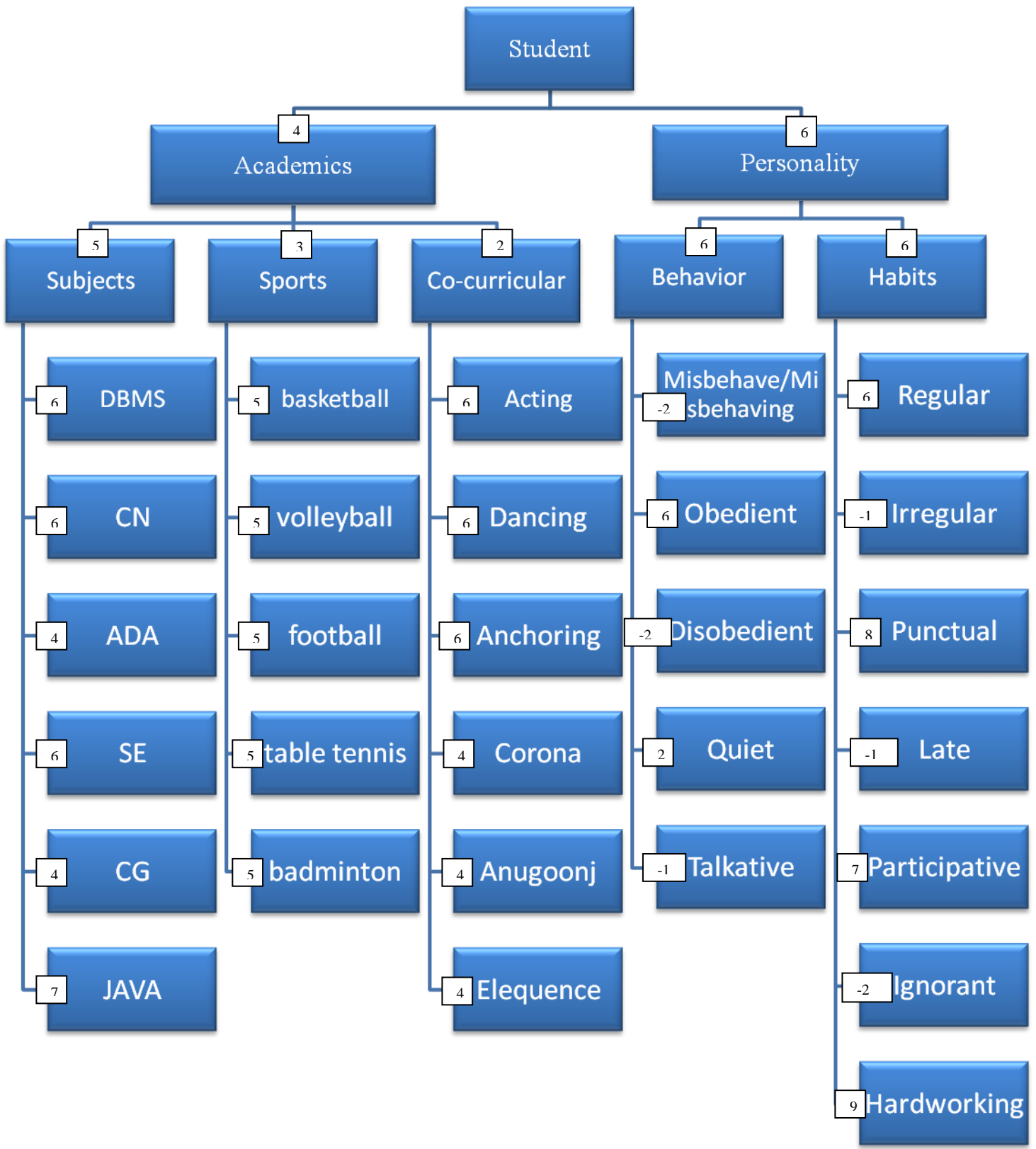

Figure 1. Aspect Tree

TABLE I
PROPOSED ALGORITHM

1. Fetch one sentence from the review.
2. Search an aspect in the sentence.(A tree of aspects is maintained)
3. When found, point to that node location of that aspect in the tree.
4. Traverse from that node to the root. Also maintain a counter 'c' that counts the number of branches in that traversal.
5. each branch is provided a specific weigh, based on the aspect to which that branch extends
6. Multiply the weighs of the branches while traversing up to the root. The counter is incremented every time a branch is traversed.
7. $$\prod_{i=node}^{i=root} w_{bi}$$

    where $w_{bi}$ is the weight of the branch $b_i$.
8. The product of weights attained when reached at the root is the aspect_value. The aspect value is accompanied by what has been said about the aspect and that is the adjectives used in that particular sentence.
9. Search for the adjectives and adverbs used for that aspect.
10. Based on the value stored in the sentiment dictionary about these adjectives and adverbs. Calculate that whether the opinion about that aspect is negative or positive. Let the calculated opinion value be op_value.
11. Now the overall sentiment of the reviewer for that particular sentence is calculated by the function $f(g,s)$

    where, $g$ is aspect
    and $s$ is sentiment about that aspect.

    $f(g,s) = (aspect\_value * op\_value)/10^{(c-1)}$

12. Now the average opinion of the review is calculated by summation of values of all the sentences in that review

    $(\sum f_i(g,s)) / n$

    where $i$ is from 1 to $n$
    and $n$ is the number of sentences in that review.

Op_value is higher for good/supporting opinion and op_value is lower for poor/opposing opinion. Aspect_value is greater if that aspect is deep down the tree that means the aspect taken up by the reviewer is more clear and more specific. The algorithm becomes clear with the case study explained below.

## IV. CASE STUDY

A case study is considered in this section where in teacher's remarks are taken. On the basis of these remarks the aspect value and the opinion value is calculated. Each sentence of the remark is analysed on the basis of the proposed tree. Following are the remarks that are under analysis.

(1) i) He is an obedient student. ii) He scored good marks in DBMS. iii) He is regular when it comes to attendance. iv) He should be more participative in co-curricular activities.

(2) i) She is a very punctual student. ii) Actively participates in co-curricular activities. iii) She is good in academics. iv) She is an elegant dancer but she is very talkative.

## V. RESULT

First sentence in the first remark is analysed and obedient word is searched in the tree and aspect value is obtained by traversing the tree which comes out to be 216. Opinion value is calculated using the dictionary based method [7]. Similarly following sentences are analysed. The summarized value is calculated using the formula in step 11 of algorithm. The average opinion for whole remark is calculated using formula in step 12 of algorithm. The results are shown in the table below.

Table 2. Case Study

| Sentence | Aspect Value(g) | Opinion Value(s) | Summarized value |
|---|---|---|---|
| i) | 216 | 7 | 15.12 |
| ii) | 120 | 6 | 7.2 |
| iii) | 216 | 9 | 19.44 |
| iv) | 8 | 4 | 3.2 |
| Average opinion for remark (1) **11.24** | | | |
| i) | 288 | 7 | 20.16 |
| ii) | 8 | 8 | 6.4 |
| iii) | 4 | 7 | 28 |
| iv) | 48 | 5 | 24 |
| Average opinion for remark (2) **19.64** | | | |

## VI. CONCLUSION

Opinion mining and sentiment analysis plays an important role in evaluating the information. In this paper we have propposed algorithm for aspect value clubbed with opinion value and sentiment value. Aspect value gives the clarity of the aspect. A lower aspect value signifies a general remark whereas a higher aspect value represents specicific remarks about an aspect. The polarity of remarks is given by the opinion value. By the proposed algorithm, we get a value that defines the aspect as well as the polarity of remarks. A clubbed value (aspect value and opinion value) plays a vital role. It gives a clear picture for analysing the opinion. The case study evaluated by the proposed algorithm proves our proposal. In future, this algorithm can be enhanced by handling dynamic aspects in the reviews and accomodating them into the aspect tree.